\documentclass[16pt,
 reprint,
 amsmath,amssymb,
 aps,
]{revtex4-1}

\usepackage{graphicx}
\usepackage{dcolumn}
\usepackage{bm}
\usepackage{geometry}
\usepackage{braket}
\usepackage{bbm}
\usepackage{lipsum}
\usepackage{mathtools}
\usepackage{lipsum}
\usepackage{xcolor}
\usepackage[colorlinks=true,allcolors=blue]{hyperref}
\newcommand*{\SavedEqref}{}
\let\SavedEqref\eqref
\renewcommand*{\eqref}[1]{%
	\begingroup
	\hypersetup{
		linkcolor=linkequation,
		linkbordercolor=linkequation,
	}%
	\SavedEqref{#1}%
	\endgroup
}

\begin{document}


\title{Construction of coherent states for Morse potential:
A su(2)-like approach}
\author{Abdessamad Belfakir}
\email{abdobelfakir01@gmail.com}
\affiliation{Equipe des Sciences de la Mati\`ere et du Rayonnement(ESMAR),\\
		Faculty of sciences,  Mohammed V University
		. Av. Ibn Battouta, B.P. 1014, Agdal,
		Rabat, Morocco.}
		\author{Evaldo M. F. Curado}
		\email{evaldo@cbpf.br}
		\affiliation{Centro Brasileiro de Pesquisas Físicas, Rua Xavier Sigaud 150, 22290-180, Rio de Janeiro, RJ, Brazil.\\ }
		\affiliation{National Institute of Science and Technology for Complex Systems, Rua Xavier Sigaud 150, 22290-180, Rio de Janeiro, RJ, Brazil.}
\author{Yassine Hassouni}
		\email{yassine.hassouni@gmail.com}
\affiliation{Equipe des Sciences de la Mati\`ere et du Rayonnement(ESMAR),\\
		Faculty of sciences,  Mohammed V University
		. Av. Ibn Battouta, B.P. 1014, Agdal,
		Rabat, Morocco.}
		


%





\begin{abstract}
\section*{abstract}
We propose a generalized su(2) algebra that perfectly describes the discrete energy part of the Morse potential. Then, we examine particular examples and the approach can be applied to any Morse oscillator and to  practically  any physical system whose spectrum is finite. Further, we construct the Klauder coherent state for Morse potential satisfying the resolution of identity with a positive measure, obtained through the solution of truncated Stieltjes moment problem. The time evolution of the uncertainty relation of the constructed coherent states is analyzed. The uncertainty relation is more localized for small  values of radius of convergence.
\end{abstract}

\pacs{Valid PACS appear here}

\maketitle

 {\quotation\noindent{\bf Keywords:}  Generalized  su(2) algebra, Coherent state, Morse potential.
  	
  	\endquotation} 
\section{Introduction}
The Morse potential \cite{Morse,Landau} is a realistic model that describes  very well the molecular vibrations inside diatomic molecules \cite{Jensen,Pauling}. Hence, it appears in the study of interactions between molecules and coherent radiations \cite{PhysRevA.19.438}. Particularly, it has been exploited in the study of the dissociation of molecules under electromagnetic fields \cite{Walker}. Since the Morse potential quantum system has paradigmatic applications in physics, chemistry \cite{Jensen,Pauling} and biology \cite{Zdravkovic}, it has been studied by using different
	approaches such as   SO(2,1) \cite{cooper1,Gerry0,Bessis,Kondo,Cizek,Huffaker,Berrondo}, SU(1,1) \cite{Balantekin,Cooper} and SU(2) groups \cite{Dong,VanIsacker,Iachello}. The latter approach has been exploited to construct the algebraic model of molecular vibrations in diatomic and Polyatomic  molecules.  More recently, It has been shown that the Morse potential can also be obtained from the Generalized Heisenberg Algebra (GHA) \cite{Monteiro1,Monteiro2,Hassouni,Angelova1}. \\
 	On the other hand, coherent states were first introduced by Schr\"{o}dinger for the quantum harmonic oscillator as some quantum states whose properties are closer to those of their classical counterpart \cite{shrodinger}. These states maintain maximum  localizability in time evolution of the system and minimize the Heisenberg uncertainty  inequality for position and momentum operators. More recently, in 1960s, Glauber \cite{Glauber} and Klauder \cite{Klauder1} have widely studied these states  in quantum optics showing their physical applicability.  Further, coherent states were not restricted only to the harmonic oscillator \cite{Gazeau} and they were constructed  for several  physical systems  such as a free particle in a square well potential \cite{Hassouni2}, Hydrogen atom \cite{Hassouni3} and P\"{o}schl-Teller potential \cite{Curado2}. These states were called nonlinear coherent states. Furthermore, it has been shown that nonlinear coherent states can be constructed and can be associated with any Lie algebra \cite{Perelomov,Gilmore,Introductory Review of Basic Notions2019} such as su(2), su(1,1) and GHA coherent states \cite{Hassouni2}. Moreover, we notice that there are several approaches to construct coherent states such as Klauder and Perelomov-Gilmore approaches \cite{Klau,Perelomov}.\\ For anaharmonic potentials, particularly for Morse potential, significant efforts have been made to construct the associated coherent states \cite{Shi,Angelova4,Roy,Maia,Daoud}. In \cite{Angelova4}, they were introduced as some superposition of the energy eigenstates which are "almost" eigenvectors of the annihilation operator. However, these states are not Klauder coherent states \cite{Klau} since they do not satisfy the resolution of unity property. Furthermore, the so-called Gazeau–Klauder coherent state for the Morse potential in \cite{Roy} satisfy the resolution of unity with a non positive weight function. Thus, they are not Klauder coherent states.\\
 	In this paper, we show that the bound states of the Morse potential can be perfectly described  by the generalized su(2). This approach is relevant because  \textit{no restriction} on the creation operator of the algebra is needed compared with su(2) and GHA approaches \cite{Dong,Angelova1} where the algebraic relation of the algebra generators are not valid for all the bound states spectrum, i.e., for all energy eigenvectors. We also aim to construct the Klauder coherent states for the bound states of Morse potential associated with  generalized su(2) algebra and discuss  their resolution of identity property.  Our approach bears some analogy with \cite{PhysRevA.64.013817} but here we use the solution of the truncated Stieltjes moment problems to construct the Klauder coherent states of Morse potential while in \cite{PhysRevA.64.013817}  the resolution of Stieltjes and Hausdorff moment problems has been used to construct coherent states for systems whose energy spectrum is infinite. To the best of our knowledge, this approach has not been exploited before and it can be used to construct Klauder coherent states for systems whose spectrum is finite. \\
 	This paper is organized as follows: in section (\ref{section2}) we show that the Morse potential can be obtained from generalized  su(2) by providing the adequate characteristic function of the algebra. Then,  in section (\ref{section3}) we shall construct the Klauder coherent state associated with Morse potential and provide the weight function satisfying the resolution of identity. Then, we  examine particular example. Moreover, in (\ref{section4}) we shall study  the properties of the constructed coherent states. Finally, our conclusions are given in (\ref{section5}).
 		\section{  Generalized su(2) and Morse potential}\label{section2}
 			\subsection{ Introduction to Generalized su(2) algebra}
  Let $J_{0}$, $J_{+}$ and $J_{-}$ be three operators satisfying the following relations \cite{Curado00,Curado01}
 	\begin{equation}\label{J-}
 	J_{0}J_-=J_-f(J_{0}),
 	\end{equation}
 	\begin{equation}\label{J+}
 	J_{+}J_{0}=f(J_{0})J_{+},
 	\end{equation}
 	and 
 	\begin{equation}\label{commu}
 	[J_{+},J_{-}]=J_{0}(J_{0}+1)-f(J_{0})(f(J_{0})+1),
 	\end{equation}
 	where $J_{0}=(J_{0})^{\dagger}$,  $(J_{+})^{\dagger}=J_{-}$ and  $f(J_{0})$ is an analytical function of $J_{0}$, called the characteristic function of the algebra. The generator $J_0$ can be any hermitian operator. The Casimir operator of the algebra defined in (\ref{J-})-(\ref{commu}), called the generalized su(2) \cite{Curado00,Curado01}, is  given by 
 	\begin{equation}\label{Casimir}
 	C=J_{+}J_{-}+f(J_{0})(f(J_{0})+1)=J_{-}J_{+}+J_{0}(J_{0}+1).
 	\end{equation}
 	By substituting the linear function $f(J_{0})=J_{0}-1$, the relations (\ref{J-})-(\ref{commu}) recover the well-known su(2) algebra \cite{Curado00,Curado01}.\\
 	Let us now provide the irreducible representation of the generalized su(2) algebra generators. Considering an eigenvector  $\left\vert n \right\rangle$  of the Hermitian  operator $J_{0}$ associated with the eigenvalue $\varepsilon_{n}$, i.e., $J_{0}\left\vert n \right\rangle=\varepsilon_{n}\left\vert n \right\rangle$ and let 
 	\begin{equation}\label{varepsilon}
 	\varepsilon_{n-1}=f(\varepsilon_{n})< \varepsilon_{n}  \quad \text{for} \quad n=0,1,\dots.
 	\end{equation}
 	The representation of the algebra is finite if and only if \cite{Curado00,Curado01}
 		\begin{equation}\label{condition1}
 	f(\varepsilon_{0})=-\varepsilon_{n_\text{max}}-1,
 	\end{equation}
  where $\varepsilon_{n_\text{max}}$, $\varepsilon_0$ are the largest eigenvalue and  the lowest eigenvalue of $J_{0}$, respectively. By using  (\ref{J-})-(\ref{commu}) and applying the Casimir operator (\ref{Casimir}) on the largest eigenvector $\ket{n_\text{max}}$ and assuming that $J_+\ket{n_\text{max}}=0$, we get
 	\begin{equation}\label{j+action}
 	J_{+}\left\vert n \right\rangle=N_{n}\left\vert n+1\right\rangle,
 	\end{equation}
 	\begin{equation}\label{j-action}
 	J_{-}\left\vert n \right\rangle=N_{n-1}\left\vert n-1\right\rangle,
 	\end{equation}
 	where
 	\begin{equation}\label{Nl2}
 	N_{n}^2=(\varepsilon_{n_\text{max}}-\varepsilon_{n})(\varepsilon_{n_\text{max}}+\varepsilon_{n}+1).
 	\end{equation}
 and $n=0,1,\dots n_\text{max}$.
 	Then, for $n=0,1,\dots n_\text{max}$, the representation is determined by giving  the largest eigenvector $\left\vert n_\text{max} \right\rangle$ associated with  the eigenvalue $\varepsilon_{n_\text{max}}$ and an eigenvector  $\left\vert n_\text{max}-l \right\rangle$ is just the action of $J_{-}$, $l$ times on $\left\vert n_\text{max} \right\rangle$  and $\varepsilon_{n_{max}-l}$ is nothing but the $l$-th iterate of $\varepsilon_{n_\text{max}}$ under $f$. The representation is then finite and the dimension of the vector space of representation is $d=n_\text{max}+1$.\\
 	We have $\varepsilon_{0}=f^{d-1}(\varepsilon_{n_\text{max}})$, implying that the condition (\ref{condition1}) can be written in the following form
 	\begin{equation}\label{condition}
 	f^d(\varepsilon_{n_{max}})=-\varepsilon_{n_\text{max}}-1,
 	\end{equation}
 		where $d$ is the dimension of the representation and $f^d(\varepsilon_{n_\text{max}})$ denotes the $d$-iterate of $\varepsilon_{n_\text{max}}$ under $f$ \cite{Curado00,Curado01}.\\
 		The characteristic function of the algebra $f$ is a decreasing function. Thus, the algebra is compatible if and only if the following condition is satisfied \cite{Curado00,Curado01} 
 	\begin{equation}\label{condition3}
 	\varepsilon_{n_{max}}\geq - \dfrac{1}{2}.
 	\end{equation}
 	\subsection{The Morse potential}
  The one-dimensional Morse model describing the vibrations of two oscillating atoms of masses $m_1$
  and $m_2$ is given by the following Schr\"{o}dinger equation \cite{Morse,Landau}
 	\begin{multline}
 	\mathcal{H}\psi(x)=\left(\frac{\hat{P}^{2}}{2m_{r}}+V(x)\right)\psi(x)\\\hspace{0.9cm}=\left(-\frac{\hbar^{2}}{2m_{r}}\dfrac{d^2}{dx^2}+V_{0}(e^{-2\beta x}-2e^{-\beta x})\right)\psi(x)\\=E\psi(x)\hspace{4.8cm},
 	\end{multline}
 	where  $ x $  is the displacement of the two atoms from their equilibrium positions. $V_{0}$
 represents the depth of the potential well at the equilibrium $x=0$. While,  $\beta$ is related to the width of the potential and $m_{r}$  is the
 	reduced mass of the oscillating system, i.e., \begin{equation}\dfrac{1}{m_r}=\dfrac{1}{m_1}+\dfrac{1}{m_2}.\end{equation}
 	The eigenvalues of the bound states are given by
 	\begin{equation}
 	E_{n}=-\frac{\hbar^{2}\beta^{2}}{2m_{r}}(p-n)^{2},
 	\end{equation}
 	where
 	\begin{equation}
 	p=\frac{\nu-1}{2},\hspace{1cm}\nu=\sqrt{\frac{8m_{r}V_{0}}{\beta^{2}\hbar^{2}}},
 	\end{equation}
 $\nu$ is related to the spectroscopic constants of  diatomic molecules \cite{Child}  and $n$ takes a finite number of values,  $\lbrace n=0, 1, 2, ..., n_{\text{max}}=[p]\rbrace$   where $[p]$ denotes the integer part of $p$ that determines the number of bound states. In general, the parameter $p$ is not an integer.\\
 	The associated eigenfunctions are given as follows
 	\begin{equation}\label{psi}
 	\psi_{n}^{\nu}(y)=\mathcal{N}_{n}e^{-\frac{y}{2}}y^{s}L_{n}^{2s}(y),
 	\end{equation}
 	where we have used the change of variable  $y=\nu e^{-\beta x}$, $L_{n}^{2s}(y)$ are the Laguerre polynomials, $2s=\nu-2n-1$ and
 	$ \mathcal{N}_{n} $ is the normalization constant given by
 	\begin{equation}\label{Norm}
 	\mathcal{N}_{n}=\sqrt{\frac{\beta(\nu-2n-1)\Gamma(n+1)}{\Gamma(\nu-n)}},
 	\end{equation}
 	where $\Gamma$ is the gamma function.
 	\subsection{Morse potential from generalized su(2)}\label{C}
 	Now, we show that the Morse potential can be described by the generalized su(2) by providing the adequate algebra generators and the appropriate characteristic function. Let $J_{0}=H+b$, where $H$ is the dimensionless Hamiltonian of the Morse potential, i.e., $H=2m_{r}\dfrac{\mathcal{H}}{\hbar^2\beta^2}$ and $b$ is a constant without dimension. We will show the reason why we choose this operator as one generator of the algebra instead of the Hamiltonian $H$. The spectrum of $J_{0}$ is given by 
 	\begin{equation}\label{varepsilonMorse}
 	\varepsilon_{n}=b-(p-n)^2,
 	\end{equation}
 from  (\ref{varepsilon}) and (\ref{varepsilonMorse})	it follows  that the characteristic function of a variable $z$ can be written as 
 	\begin{equation}\label{functionMorse}
 	f(z)=z-2\sqrt{b-z}-1.
 	\end{equation}
Now, let us provide the constant  $b$ satisfying the cut condition (\ref{condition1}). Remembering that $p$ is not an integer in general, it can always be written as $p=n_{max}+x$ where $x\in]0,1[$. By substituting  (\ref{varepsilonMorse}) and (\ref{functionMorse}) in (\ref{condition1}), the cut condition  (\ref{condition1}) can now be written as 
\begin{equation}
2 b-(p+1)^2-x^2+1=0.
\end{equation}
By solving this equation, we find that 
\begin{equation}\label{b}
b=\frac{1}{2} \left(p^2+2 p+x^2\right).
\end{equation}
Consequently, the spectrum (\ref{varepsilonMorse}) can be simplified as
\begin{equation}\label{varepsilonx}
\varepsilon_{n}=\frac{1}{2} \left(-2 (n-p)^2+p^2+2 p+x^2\right).
\end{equation}
By replacing $p$ by $n_{max}+x$, the eigenvalue $\varepsilon_{n}$ can be written in terms of $n$, $n_{max}$ and $x$ as
\begin{equation}\label{n}
\varepsilon_{n}=-n^2+n_\text{max} (2 n-x+1)+2 n x-\frac{n_\text{max}^2}{2}+x.
\end{equation}
It follows that the lowest eigenvalue can be given by 
\begin{equation}
\varepsilon_{0}=n_\text{max}(-\frac{n_\text{max}}{2}+ 1-x)+x,
\end{equation}
and the largest eigenvalue can now be  written as
\begin{equation}\label{largest}
\varepsilon_{n_{max}}=n_\text{max}(\frac{n_\text{max}}{2}+ x+1)+x.
\end{equation}
Thus, the cut condition for the representation to be finite is verified. From (\ref{n}), we find 
\begin{equation}\label{-1}
\varepsilon_{-1}=f(\varepsilon_{0})=-\frac{1}{2} n_\text{max} (n_\text{max}+2 x+2)-x-1.
\end{equation}
Consequently, from (\ref{n}) and (\ref{-1}), the cut condition $f(\varepsilon_{0})+\varepsilon_{n_\text{max}}+1=0$ is verified. Then, the Morse potential can be totally described  by the generalized su(2). This result is valid for any Morse oscillator, i.e., any value of $p$.\\
By considering $J_0=H$, i.e., $b=0$, the cut condition for the representation to be finite  $f(\varepsilon_{0})+\varepsilon_{n_\text{max}}+1=0$ does not admit any solution showing the importance of the added constant $b$.\\
The representation of the algebra generators can be easily obtained. From (\ref{Nl2}), (\ref{n}) and (\ref{largest}) we show that 
\begin{widetext}
\begin{equation}\label{Nn}
N_n^2=(n+1) (n_\text{max}-n) (n_\text{max}-n+2 x) (2 n_\text{max}-n+2x+1).
\end{equation}
\end{widetext}
It is easily seen that the quantity $N_n^2 $ defined in (\ref{Nn}) is positive. Then, its square root is well defined and the action of $J_+$ and $J_-$ can be easily obtained. Thence, it is easy to check that the cut conditions $N_{-1}=0$  and $N_{n_\text{max}}=0$ which imply that $J_-\ket{0}=0$ and $J_+\ket{n_\text{max}}=0$ are verified.
 	\subsection{Particular examples}
 	Now, let us apply the results above to particular Morse oscillators. 
 	\subsubsection{A Morse potential with 3 levels}
 	In fact, this Morse oscillator is not related to any physical system but we examine it to show how the Morse potential can be described by generalized su(2). However, the approach  can be used for any Morse oscillator. From (\ref{n}) with $p=2+x$,  $x\in ]0,1[$,  we have in this case 
 	\begin{equation}
 	\varepsilon_{0}=-x,
 	\end{equation}
 	\begin{equation}
 	\varepsilon_{1}=x+3,
 	\end{equation}
 	and
 	\begin{equation}
 	\varepsilon_{2}=3x+4.
 	\end{equation}
 From (\ref{Nn}), we have $N_{-1}=0$, $N_2=0$ and 
 \begin{equation}
N_0^2= 4 ( x+1) ( 2 x+5),
  \end{equation}
 \begin{equation}
N_1^2=4 ( x+2) ( 2 x+1),
 \end{equation}
Then, the action of $J_+$ and $J_-$ on $\ket{n}$ with $n=0,1,2$ are well defined and $J_+\ket{n_\text{max}}=J_-\ket{0}=0$ for $x\in ]0,1[$.
\subsubsection{$n_\text{max}=7$}\label{D}
Let us consider as a theoretical example the Morse oscillator with $p=7.7$. It follows that $x=0.7$ and by using (\ref{b}) we find that the adequate value of $b$ for which this system can be described by the algebra is $b=37.59$, and the action of $J_+$ and $J_-$ on $\ket{n}$ for $n=0,1,\dots,7$ can be easily  found by using (\ref{Nn}). We can see that $N_{-1}=N_{7}=0$ which means that the cut conditions on the largest and the lowest eigenvectors are satisfied. 
 		\section{Construction of coherent states for Morse potential}\label{section3}
 In this section, we construct the generalized su(2)  coherent state associated with Morse potential. let us consider a vector $\ket{z}$ defined by
 	\begin{equation}\label{CS}
 	\left\vert z \right\rangle=\textbf{N}(|z|)\text{e}^{z J_{+}}\left\vert  0 \right\rangle,
 	\end{equation}
 	where $\textbf{N}(|z|)$ is the normalization function, $z$ is a complex number and $e^x$ is the  usual exponential function. From(\ref{j+action}), We have
 	\begin{equation}
 	J_{+}^n\ket{0}=\prod_{i=0}^{n-1}N_{i}\ket{n},
 	\end{equation}
 	and 
 		\begin{equation}
 	J_{+}^{n_\text{max}+1}\ket{0}=0,
 	\end{equation}
 	Thus, the  state $\ket{z}$ reads 
 	
 	\begin{multline}\label{CS2}
 	\left\vert z \right\rangle=\textbf{N}(|z|)\sum_{n=0}^{n_{max}}\dfrac{z^{n}J_{+}^n}{n!}\left\vert 0 \right\rangle\\=\textbf{N}(|z|)\sum_{n=0}^{n_{max}}\dfrac{z^{n}\prod_{i=0}^{n-1}N_{i}}{n!}\left\vert n\right\rangle.\hspace{2.3cm}
 	\end{multline}
 We note  $N_{n-1}!=\prod_{i=0}^{n-1}N_{i}$, and by definition $N_{-1}!:=1$. Then, the state (\ref{CS2}) can be written as 
 \begin{equation}\label{CS3}
\ket{z} =\textbf{N}(|z|)\sum_{n=0}^{n_{max}}\dfrac{z^{n}N_{n-1}!}{n!}\left\vert n\right\rangle.
 \end{equation} It is said that the state $\left\vert z\right\rangle$ is a Klauder coherent state if and only if it satisfies the following conditions\\ 
 	i)normalization $\braket{z|z}=1$ , \\
 	ii) continuity in the label,
 	\begin{equation}
 	|\left\vert z \right\rangle - \left\vert z' \right\rangle|\longrightarrow 0 \quad \text{when } \quad | z- z'|\longrightarrow 0
 	\end{equation}
 	iii) completeness or resolution of identity
 	\begin{widetext} \begin{equation}\label{completeness}
 	\int\int d^2z \dfrac{w(|z|^2)}{\pi}\left\vert z \right\rangle\left\langle z \right\vert=\sum_{n=0}^{n_{max}}\ket{n}\bra{n}=\mathbbm{1},
 	\end{equation}
 \end{widetext}
 	where $\mathbbm{1}$ denotes the identity operator of the vector space of representation, and $w(|z|^2)$  is a positive function called the weight function \cite{PhysRevA.64.013817}.\\ The summation in (\ref{CS3}) is finite. Then, the conditions (i) and (ii) are obviously satisfied. Now, let us examine the condition (iii). Let $z=re^{i\theta}$ where $0\leq r<\infty$ and $0\leq \theta\leq 2\pi$. Then, $d^2z=rdrd\theta$. By substituting (\ref{CS3}) in (\ref{completeness}), we find that 
 	\begin{widetext}
 	\begin{equation}
 	\int_{0}^{\infty}\int_{0}^{2\pi} rdrd\theta \dfrac{w(r^2)}{\pi}(\textbf{N}(r^2))^2\times\sum_{n,m=0}^{n_{max},n_{max}}\dfrac{r^{n+m}\text{e}^{i(n-m)\theta}N_{n-1}!N_{m-1}!}{n!m!}
 	\ket{n}\bra{m}=\mathbbm{1}.	\end{equation}
 \end{widetext}
 	We have $\int_{0}^{2\pi}d\theta\text{e}^{i(n-m)\theta}=2\pi\delta_{n,m}$. Thus,
 	\begin{equation}
 	2 \sum_{n=0}^{n_{max}}\ket{n}\bra{n}\int_{0}^{\infty}rdr\hspace{0.05cm}w(r^2)(\textbf{N}(r^2))^2\dfrac{r^{2n}(N_{n-1}!)^2}{(n!)^2}=\mathbbm{1}
 	\end{equation}
 	Let $x=r^2$. Then
 	\begin{equation}
 \sum_{n=0}^{n_{max}}\ket{n}\bra{n}\int_{0}^{\infty}dx\hspace{0.05cm}w(x)\left(\textbf{N}(x)\right)^2\dfrac{x^{n}(N_{n-1}!)^2}{(n!)^2}=\mathbbm{1}.\end{equation}
 	Let $h(x)=(\textbf{N}(x))^2w(x)$. Finally, the completeness problem reduces to the resolution of the following moment problem
 	\begin{equation}\label{mom}
 	\int_{0}^\infty dx h(x) x^n=\dfrac{(n!)^2}{(N_{n-1}!)^2},\quad n=0,1,\dots ,n_{max}.
 	\end{equation}
  Let 
 	\begin{equation}\label{mat}
 	d\mu(x)=h(x)dx
 	\end{equation} 
 	and let $s_n=\dfrac{(n!)^2}{(N_{n-1}!)^2}$. It follows then that 
 	\begin{equation}\label{moment}
 	\int_{0}^{\infty}d\mu(x)x^n=s_{n},\quad n=0,1,\dots,n_{max}.
 	\end{equation}
  The moment problem (\ref{moment}) is known by the  truncated Stieltjes moment problem \cite{lawrence}.\\
 	Now, let us recall the condition under which (\ref{moment}) admits a solution \cite{lawrence}. 
 	Let 
 	\begin{equation}\label{A(k)}
 	A(k)=\begin{pmatrix} 
 	s_{0}& s_{1}& \dots & s_{k} \\
 	s_{1}& s_{2}& \dots & s_{k+1} \\ \vdots& \vdots & \ddots & \vdots \\ s_{k}& s_{k+1}& \dots & s_{2k} 
 	\end{pmatrix},
 	\end{equation}
 	\begin{equation}\label{B(k)}
 	B(k)=\begin{pmatrix} 
 	s_{1}& s_{2}& \dots & s_{k+1} \\
 	s_{2}& s_{3}& \dots & s_{k+2} \\ \vdots& \vdots & \ddots & \vdots \\ s_{k+1}& s_{k+2}& \dots & s_{2k+1}
 	\end{pmatrix},
 	\end{equation}
 	and \begin{equation}
 	v(k+1,k)=\begin{pmatrix} 
 	s_{k+1} \\
 	s_{k+2} \\ \vdots \\  s_{2k+1} 
 	\end{pmatrix}.
 	\end{equation}
 	For $n_{max}=2k+1$, if $A(k)\geq 0$, $B(k)\geq0$ i.e., $A(k)$ and $B(k)$ are positive  semi-definite matrices and $v(k+1,k)\in \text{Rank} (A(k))$, then there exists a measure $d\mu(x)$ satisfying (\ref{moment}), implying that the state $\left\vert z \right\rangle$ is Klauder coherent state. Let us note that if $A(k)$ and $B(k)$ are positive definite matrices, the third condition is obviously satisfied, i.e.,  $v(k+1,k)\in Rank (A(k))$ is obviously satisfied \cite{lawrence}.\\
 	For $n_{max}=2k$, there exists a measure $d\mu(x)$ satisfying (\ref{moment}), i.e., $\left\vert z \right\rangle$ is Klauder coherent state, if and only if  $A(k)\geq 0$, $B(k-1)\geq0$ and $v(k+1,k-1)\in \text{Rank} (B(k-1))$. If $A(k)>0$ and $B(k-1)>0$, it follows that  $v(k+1,k-1)\in \text{Rank} (B(k-1))$ is satisfied \cite{lawrence}. Thus, we have provided the conditions under which the vector $\ket{z}$ defined in (\ref{CS3}) is a Klauder coherent state. If these conditions are not satisfied, the vector $\ket{z}$ is just a normalized state but never a coherent state as it does not  satisfy the completeness condition. The form of the measure satisfying the moment (\ref{moment}) is given by \cite{lawrence}
 	 \begin{equation}\label{matt}
 	\mu(x)=\sum_{j=0}^{k}\rho_j\delta(x-y_j),
 	\end{equation} 
 	 where $\rho_j\geq0$, $y_j\in \mathbbm{R}$ and $\delta(x-y_j)$ denotes the Dirac delta function. Thus, compared with coherent states constructed in \cite{Shi,Angelova4,Roy,Maia,Daoud}, we have provided here a construction of coherent states for the bound states of the Morse potential from the solution of the truncated Stieltjes moment problem and provided the associated positive weight function.\\
 	For example, the Gazeau-Klauder coherent states of Morse potential constructed in \cite{Roy} are given by 
 	\begin{equation}\label{zgamma}
\ket{z,\gamma}=N(|z|)\sum_{n=0}^M\dfrac{z^{n/2}e^{-i\gamma e_n}}{\sqrt{\rho_n}}\ket{n},
 	\end{equation}
 	 where $z\geq 0$, $-\infty<\gamma<\infty$, $e_n=(M+1)^2-(M+1-n)^2$ and $M$ is a positive integer. The normalization  function $N(|z|)$ now reads
 	 \begin{equation}
N(|z|)=\left(\sum_{n=0}^M\dfrac{z^{n}}{\rho_n}\right)^{-1/2},
 	 \end{equation}
 	 with $\rho_n=\prod_{i=1}^n e_i$ and $\rho_0$=1. The completeness condition implies that the sequences $\rho_n$ satisfy the following integral equation
 	 \begin{equation}\label{khayb}
\int_{0}^\infty x^n h(x)dx=\rho_n.
 	 \end{equation}
 	 The function $h(x)$ satisfying (\ref{khayb}) is given by
 	 \begin{equation}
h(x)=\Gamma(2M+2)x^{-(M+1)}J_{2M+2}(2\sqrt{x}),
 	 \end{equation}
 	 where $J_n(x)$ denotes  the Bessel function of the first kind which is not positive for any $0\leq x<\infty$. Thus, for a fixed value of $M$ wich determines now the number of bound states, $h(x)$ is not a positive function. This implies that the coherent states $\ket{z,\gamma}$ constructed in \cite{Roy} and recalled in (\ref{zgamma}) do not satisfy the resolution of identity with a positive \textit{weight } function. The physical properties of the coherent states (\ref{zgamma}) have been investigated in \cite{Roy}.\\
 	 Here, we have provided an approach to construct the coherent states for the bound states of the Morse potential associated with the generalized su(2) algebra and shown that their resolution of identity can be solved with a positive measure by using the solutions of truncated Stieltjes moment problem.  Now, let us examine a particular example.
 	\subsection*{$p$=7.7}
We have shown  in (\ref{D}) that the Morse oscillator with $p=7.7$ can be totally described by generalized su(2) algebra with $b=37.59$. In this case, we have $n_\text{max}=7$, $k=3$ and  $N_0=31.0535$, $N_1= 36.98$, $N_2=37.1806$, $N_3=34.0259$, $N_4=28.6077$,  $N_5=21.5666$, $N_6=13.2182$ and $N_7=0$. The matrices $A(3)$ and $B(3)$ are  positive definite matrices. Since $A(3)>0$. Then, $v(4,3) \in \text{Rank}(A(3))$ is verified. Then, it  exists a positive measure  satisfying (\ref{moment}). Consequently, the vector $\ket{z}$ defined in (\ref{CS3}) with $p=7.7$, is a Klauder coherent state. A measure   of the form (\ref{matt}) satisfying  (\ref{moment}) with $p=7.7$ can be easily obtained.  From (\ref{moment}) and (\ref{matt})  we have
 \begin{equation}
 \sum_{j=0}^{3} \rho_j\int_{0}^{\infty}dx\delta(x-y_j)x^n=s_{n}\quad\quad n=0,1,\dots,7.
 \end{equation} 
 Then, we have
 \begin{equation}\label{u}
 \sum_{j=0}^{3} \rho_j y_j^n=s_{n},\quad\quad n=0,1,\dots,7.
 \end{equation} 
 Thus, a finitely atomic positive measure on $\mathbbm{R}$  with atoms $\rho_j$ can be constructed (see \cite{lawrence}).
 	\section{Properties of the Morse coherent state}\label{section4}
 	\subsection{The uncertainty relation on the Morse coherent state}
 	In this section, we shall study the properties of a particular Morse coherent state proposed in (\ref{CS3}). We will examine the behavior of the  uncertainties of both  position and momentum operators in terms of  $|z|$. Then, we study the behavior of the uncertainty relation in terms of $|z|$ and its time evolution for particular values of $|z|$. The dispersions of the position and the momentum on the coherent state $\ket{z}$ are given by 
 	\begin{equation}\label{desp1}
\Delta x=\sqrt{\braket{x^2}-(\braket{x})^2},
 	\end{equation}
 	and
 		\begin{equation}\label{desp2}
 	\Delta p=\sqrt{\braket{p^2}-(\braket{p})^2},
 	\end{equation}
 	respectively, where $\braket{.}=\bra{z}.\ket{z}$. Thus, The uncertainty relation is given by 
 	\begin{equation}\label{uncer}
\Delta x\Delta p=\sqrt{(\braket{x^2}-(\braket{x})^2)(\braket{p^2}-(\braket{p})^2)}.
 	\end{equation}
 	The matrices elements of the operators $x$, $x^2$, $p$ and $p^2$, in the basis spanned by the eigenvectors defined in (\ref{psi}) with $\beta=1$ are given by \cite{PhysRevA.21.1829,Angelova4}
\begin{widetext}
\begin{multline}
\bra{m}x\ket{n}=\ln (\nu )\delta
_{m,n}+\mathcal{N}_{m} \mathcal{N}_{n} \sum _{i=0}^m \sum _{j=0}^n
\frac{(-1)^{i+j+1}}{i! j!}\binom{m+(\nu-2m-1)}{m-i} \binom{n+(\nu-2n-1)}{n-j}\\\times \Gamma (\nu+i+j-m-n -1)\psi ^{(0)}(\nu+i+j-m-n -1),\hspace{2.8cm}
\end{multline}
	\begin{multline}
	\bra{m}x^2\ket{n}=\mathcal{N}_{m} \mathcal{N}_{n} \sum _{i=0}^m \sum _{j=0}^n
	\frac{(-1)^{i+j}}{i! j!}\binom{m+(\nu-2m-1)}{m-i} \binom{n+(\nu-2n-1)}{n-j}\\\times \Gamma (\nu+i+j-m-n -1)\times\{[\psi ^{(0)}(\nu+i+j-m-n -1)-\ln(\nu)]^2+\psi ^{(1)}(\nu+i+j-m-n -1)\},
	\end{multline}
\begin{equation}
\bra{n+k}p\ket{n}=i\hbar(-1)^{k+1}\mathcal{N}_{n+k} \mathcal{N}_{n}\dfrac{\Gamma(\nu-k-n)}{2n!}(1- \delta
_{k,0}),
\end{equation}
\begin{equation}
\bra{n+k}p^2\ket{n}=\hbar^2(-1)^{k+1}\mathcal{N}_{n+k} \mathcal{N}_{n}\dfrac{\Gamma(\nu-k-n)}{4n!}((k-1)\nu-k(k+2n+1)),\quad k\ne0,
\end{equation}
and 
\begin{equation}
\bra{n}p^2\ket{n}=-\dfrac{\hbar^2(2n+1)(2n+1-\nu)}{4},
\end{equation}
\end{widetext}
where $\psi ^{(1)}$ denotes the first derivative of the digamma function $\psi ^{(0)}$ and $\mathcal{N}_n$ is the normalization function given in (\ref{Norm}). Thus, the dispersions of the position and the momentum  (\ref{desp1})-(\ref{desp2}) can be computed for the coherent state (\ref{CS3}) and the uncertainty relation (\ref{uncer}) can be easily obtained.  The time evolution of the coherent state $\ket{z}$ given in (\ref{CS3}) can be obtained by the action of the evolution operator 
\begin{equation}
   \ket{z,t}= U(t)\ket{z}=\text{e}^{-i\mathcal{H}t/\hbar}\ket{z}.
\end{equation}
Thus, the time evolution of the uncertainty relation can also be easily computed.
\subsection*{Application to the case of the molecule $\text{Hg}^2\text{H}$} 
\begin{figure}[h]
 	\centering
 	\includegraphics[width=7cm]{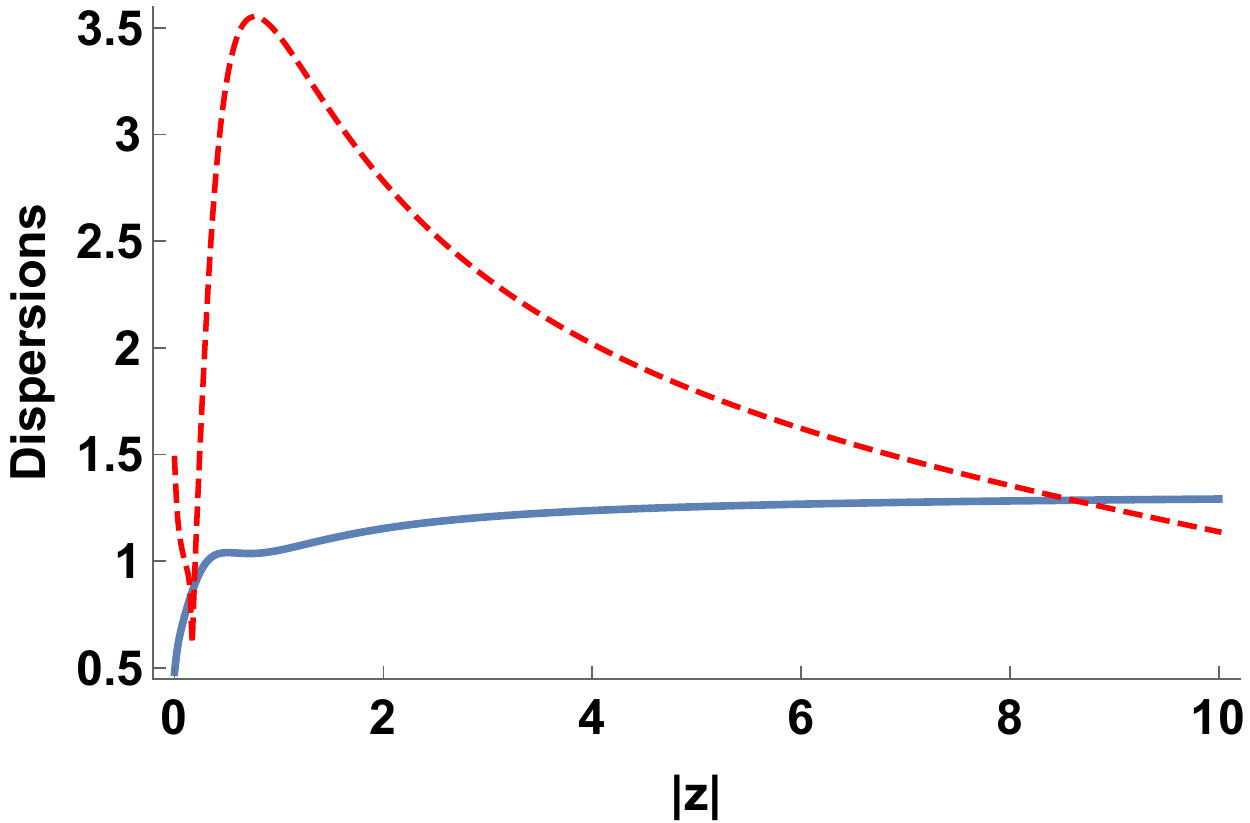}
 	\caption{The dispersions $\Delta x$ (continuous, blue curve) and  $\Delta  p_x/\hbar$ (dashed, red curve) in terms of $|z|$ for Morse coherent state with $p=9.46$.} 
 	\label{fig4}
 \end{figure}

\begin{figure}[b]
	\centering
	\includegraphics[width=7cm]{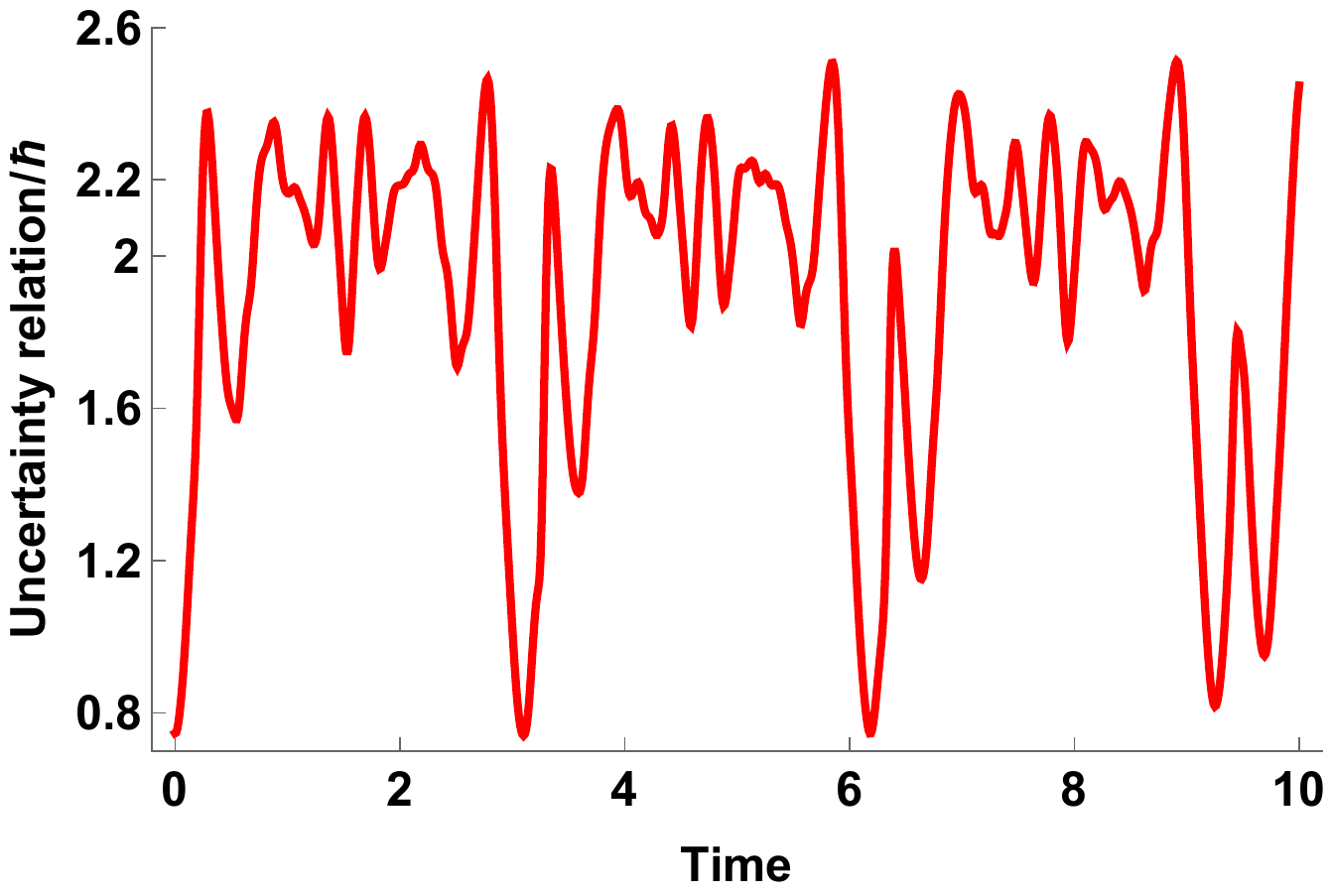}
	\caption{The behavior of time evolution of the uncertainty relation of Morse coherent state with $p=9.46$ and $|z|=0.1$} 
	\label{fig2}
\end{figure}

\begin{figure}[h]
 	\centering
 	\includegraphics[width=7cm]{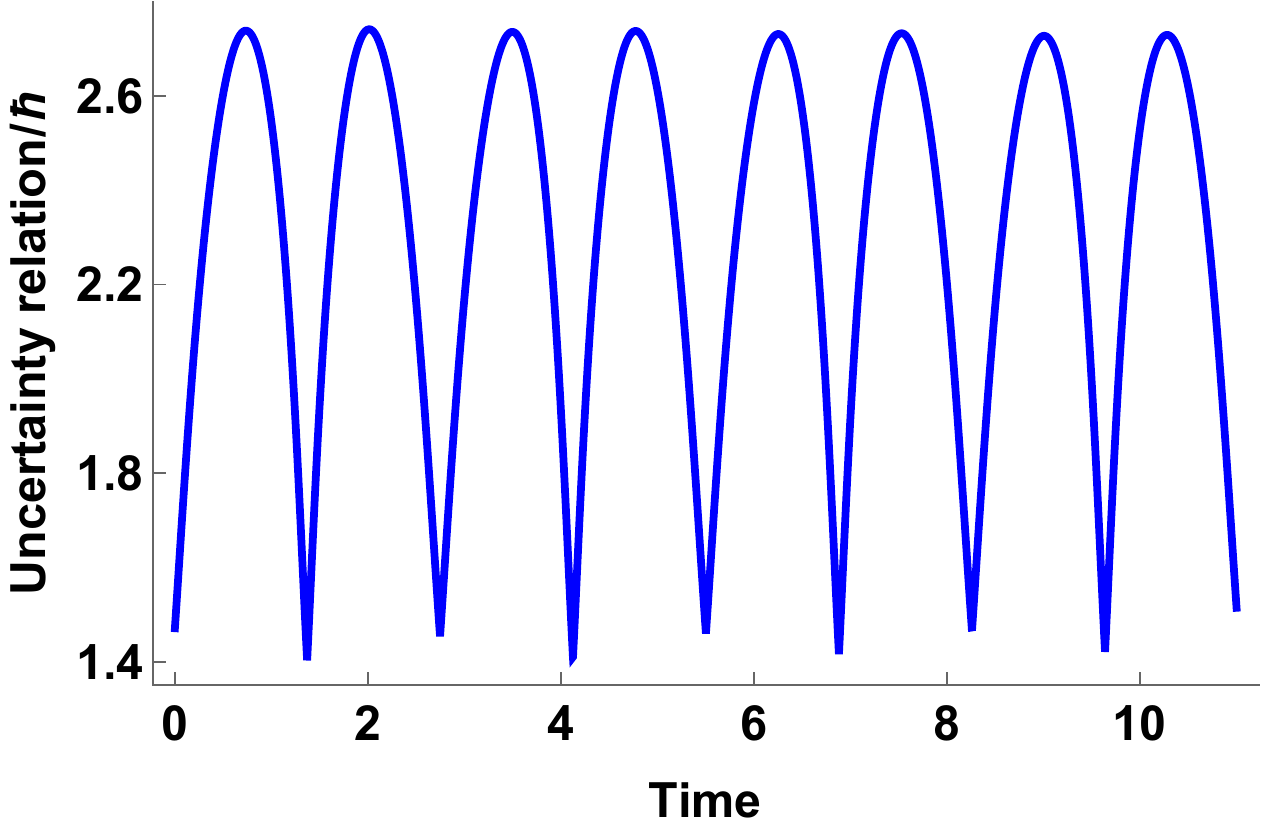}
 	\caption{The behavior of time evolution of the uncertainty relation on Morse coherent state with $p=9.46$ and $|z|=10$.} 
 	\label{fig3}
 \end{figure}

Now, we apply the results above to the molecule $\text{Hg}^2\text{H}$. The parameter $\nu$ is related to the experimentally measured molecular harmonicity $\omega_e$ and the anharmonicity $\omega_e x_e$ constants by 
\begin{equation}
\nu=\dfrac{\omega_e}{\omega_e x_e}.
\end{equation}
By using the results given in \cite{Herzberg}, for the state $ \text{X}^2\Sigma$, we have $\nu\approx 19.93$, $p=9.46$ and $n_{\text{max}}=9$. From (\ref{b}) we find that the constant $b$ associated with the characteristic function of the generalized su(2) is $b=54.31$.
Now, we shall examine the behavior of the  dispersions $\Delta x$, $ \Delta p_x$  of the generalized su(2) coherent state associated with the molecule $\text{Hg}^2\text{H}$ in terms of $|z|$. The behaviors of the dispersions $\Delta x$, $ \Delta p_x$ are shown in Figure (\ref{fig4}). Analyzing the figure (\ref{fig4}), we can see that the product $\Delta x \Delta p_x$ approaches to $0.5\hbar$ for small values of $|z|$ which means that the uncertainty relation $\Delta x\Delta p_x$ is minimized for small values of $|z|$. In Figures (\ref{fig2})-(\ref{fig3}), we show the time evolution of the uncertainty relation (\ref{uncer}) for the generalized su(2) coherent state $\ket{z,t}$ associated with the molecule   $\text{Hg}^2\text{H}$ with $p=9.46$ for $|z|=0.1$ and $|z|=10$, respectively.  In  all numerical  simulations  we  have taken $\beta=1$.
  Analyzing these Figures, we can see that the uncertainty relation oscillates between  minimum values and maximum values and that the uncertainty relation approaches to $0.5\hbar$ when  $|z|$ is very small. Consequently, the uncertainty is more localized for coherent states with small values of $|z|$.

	\begin{figure}[b]
	\centering
	\includegraphics[width=7cm]{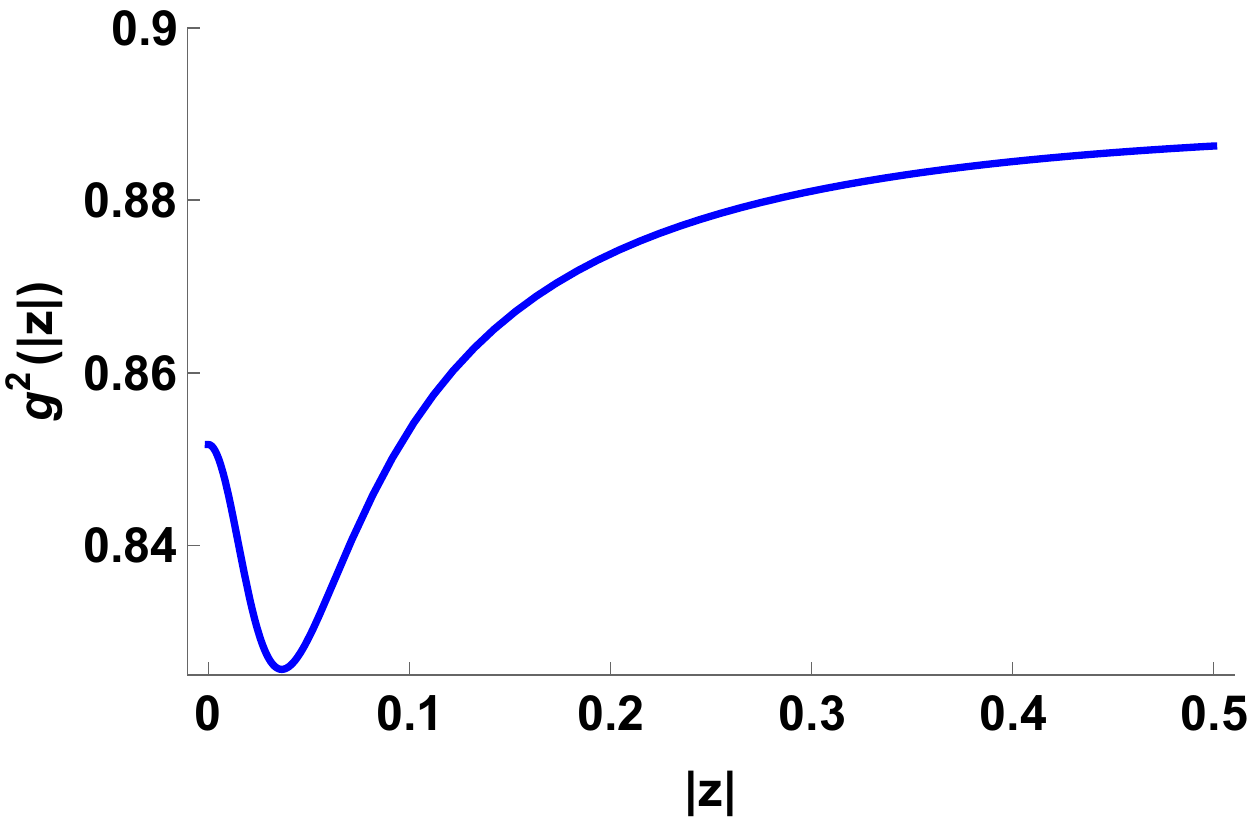}
	\caption{The behavior of $g^2(|z|)$ on Morse coherent state with $p=9.46$.} 
	\label{fig5}
\end{figure}
 	\subsection{Statistical properties of Morse coherent state}
\begin{figure}[h]
 	\centering
 	\includegraphics[width=7cm]{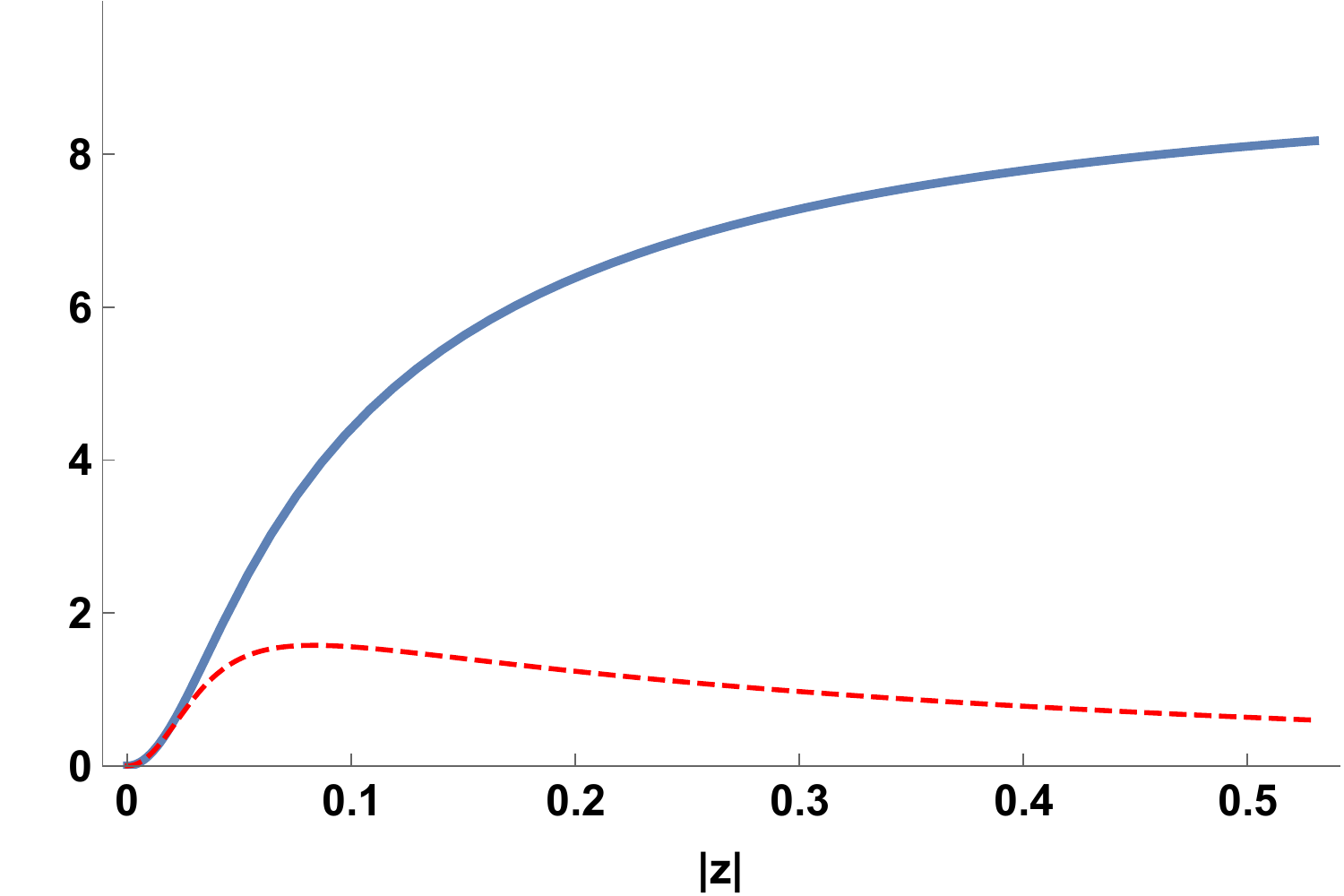}
 	\caption{The mean value of the number of particles operator $\hat{n}$ (continuous, blue curve) and its variance $(\Delta \hat{n})^2$ (dashed, red curve) in terms of $|z|$ for Morse coherent state with $p=9.46$.} 
 	\label{fig6}
 \end{figure}

 	The normalized second order correlation function at zero time on a state $\ket{\psi}$ is defined by 
 	\begin{equation}
g^{2}=\dfrac{\bra{\psi} \hat{n}^2\ket{\psi}-\bra{\psi} \hat{n}\ket{\psi}}{(\bra{\psi} \hat{n}\ket{\psi})^2},
 	\end{equation}
 	where $\hat{n}$ is the particle number operator. In quantum optics, the function $g^{2}$ characterizes the non-classical nature of the light field described by the state $\ket{\psi}$ \cite{Glauber1}. The light is antibunched  if $g^{2}<1$, coherent if $g^{2}$=1, bunched if $1<g^{2}<2$ and thermal when $g^{2}=2$. We note that several measures of nonclassicality of light fields have been proposed such as the Wigner function \cite{wignerclass}, the Mandel parameter \cite{Mandel} and the squeezing  parameter \cite{Yuen}.\\
 	For the generalized su(2) Morse coherent state $\ket{z}$ associated with the molecule $\text{Hg}^2\text{H}$ with $p=9.46$, the second order correlation function 
 	\begin{equation}\label{secondorderz}
 	g^{2}(|z|)=\dfrac{\bra{z} \hat{n}^2\ket{z}-\bra{z} \hat{n}\ket{z}}{(\bra{z} \hat{n}\ket{z})^2},
 	\end{equation}
 	can be easily computed by using the fact that 
 	\begin{equation}\label{n2z}
\bra{z}\hat{n}^2\ket{z} =(\textbf{N}(|z|))^2\sum_{n=0}^{n_{max}}\dfrac{n^2\hspace{0.05cm} |z|^{2n}(N_{n-1}!)^2}{(n!)^2},
 	\end{equation}
 	and 
 		\begin{equation}\label{nz}
 	\bra{z}\hat{n}\ket{z} =(\textbf{N}(|z|))^2\sum_{n=0}^{n_{max}}\dfrac{n \hspace{0.05cm}|z|^{2n}(N_{n-1}!)^2}{(n!)^2}.
 	\end{equation}
 The behavior of $g^2(|z|)$ given in (\ref{secondorderz}) for the Morse coherent state with $p=9.46$, $b=54.31$ and $n_{\text{max}}=9$ is shown in the figure (\ref{fig5}). Immediately, we see that $0<g^2(|z|)<1$. This shows the  bunching effects of the Morse coherent states $\ket{z}$ associated with the molecule $\text{Hg}^2\text{H}$. The calculations of the second order correlation function of Morse coherent states for other diatomic molecules can be obtained in a similar way.\\
 By using (\ref{n2z}) and (\ref{nz}), the variance of the number operator $(\Delta\hat{n})^2=\bra{z}\hat{n}^2\ket{z}-(\bra{z}\hat{n}\ket{z})^2$ can be easily computed. In figure (\ref{fig6}), we show the behaviors of the mean value of the number of particles operator $\hat{n}$ and its variance on the coherent state $\ket{z}$ associated with $p=9.46$ and  $b=54.31$ in terms of the amplitude $|z|$. Interestingly, the mean value of the operator $\hat{n}$ is shown to be larger  and the associated variance is shown to be smaller for high values of $|z|$. 
\section{FINAL COMMENTS}\label{section5}
 In this paper, we have shown that the discrete energy
 part of the Morse potential can be perfectly  described  by  the generalized su(2) algebra.
 By using this approach no condition on the action of ladder operators is needed compared with  su(2) and  GHA approaches. Once the characteristic function of the algebra is determined, the algebraic relations are valid for all energy eigenvectors which is not the case for previous  approaches.
 We have examined particular examples of Morse systems and the approach is valid for any Morse oscillator. Then, we have used the analytical solutions of truncated  Stieltjes moment problem to construct the Klauder coherent states associated with Morse oscillator. These states satisfy the resolution of identity with a positive measure compared with those introduced in \cite{Maia, Roy}. Finally, we have analyzed the behaviors of the the time evolution of the uncertainty relation of the constructed coherent states for particular Morse oscillators in terms of $|z|$ and have shown that uncertainty approaches to $0.5\hbar$ for small values of $|z|$.
 \section*{Acknowledgement} 
One of us, E.M.F.C., would like to thank  CNPq, FAPERJ and CAPES for  financial support. This work is partially supported by the ICTP through AF-14.

\end{document}